\documentstyle[seceq,preprint,wrapfig,epsf]{ptptex}

\newcommand{\gsim}%
{\mathrel{\mbox{\raisebox{-1.0ex}
    {$\stackrel{\displaystyle >}{\displaystyle \sim}$}}}}
\newcommand{\lsim}%
{\mathrel{\mbox{\raisebox{-1.0ex}
    {$\stackrel{\displaystyle <}{\displaystyle \sim}$}}}}
\newcommand{\ie}{{\it i.e.\/} }
\newcommand{\etal}{{\it et al.\/} }
\newcommand{\vev}[1]{\left\langle #1 \right\rangle}
\newcommand{\ol}[1]{\overline{#1}}
\newcommand{\bsg}{$b\rightarrow s\gamma$ }
\newcommand{\mbsg}{\ b\rightarrow s\gamma\ }
\newcommand{\br}{{\rm Br}}

\newcommand{\Journal}[4]{{#1} {\bf #2} {(#3)} {#4}}

\newcommand{\plb}{\sl Phys.~Lett.~{\bf B}}
\newcommand{\prp}{\sl Phys.~Rep.}

\newcommand{\prd}{\sl Phys.~Rev.~{\bf D}}
\newcommand{\prl}{\sl Phys.~Rev.~Lett.}

\newcommand{\npb}{\sl Nucl.~Phys.~{\bf B}}
\newcommand{\ptp}{\sl Prog.~Theor.~Phys.}

\newcommand{\zpc}{\sl Z.~Phys.~{\bf C}}

\notypesetlogo  

\preprintnumber{
TU-494 \\
KEK-TH-468 \\
KEK Preprint 95-199
}

\title{%
Charged Higgs mass bound from the \bsg process in the minimal
supergravity%
\footnote{%
Talk given
by Toru Goto
at Yukawa International Seminar '95: From the Standard Model to Grand
Unified Theories, Kyoto, Japan, 21-25 Aug 1995.
}
}

\author{%
Toru {\sc Goto}$^{1)}$%
\footnote{%
E-mail address: \verb:goto@tuhep.phys.tohoku.ac.jp:}
and
Yasuhiro {\sc Okada}$^{2)}$\footnote{%
E-mail address: \verb:okaday@theory.kek.jp:}
}

\inst{%
$^{1)}$Department of Physics, Tohoku University, Sendai 980-77 Japan
\\
$^{2)}$Theory Group, KEK, Tsukuba, Ibaraki, 305 Japan
}

\recdate{%
\today
}

\abst{%
The constraint on the mass of the charged Higgs boson from the recent
measurement of the inclusive \bsg decay is studied in a framework of the
minimal supergravity.
It is shown that the lower bound for the charged Higgs mass crucially
depends on the sign of the higgsino mass parameter ($\mu$).
For $\mu<0$, the bound exceeds 400 GeV when the ratio of two Higgs
vacuum expectation values ($\tan\beta$) is larger than 10.
No strong bound is obtained for $\mu>0$ due to cancellations between
charged Higgs and supersymmetric particle contributions.
}

\begin{document}

\maketitle

\section{Introduction}

Flavor changing neutral current (FCNC) processes play a unique role 
in searching for physics beyond the standard model (SM) of 
elementary particles.  
These processes are sensitive to virtual effects of new particles, 
since the FCNC processes in SM do not occur at the tree level. 
These processes can thus be more powerful than direct particle searches 
in putting constraints on the parameter space of various new physics. 
In particular, the radiative decay of the $b$ quark, \bsg, deserves 
a special attention.  
Recently, the CLEO group \cite{cleo} has reported the first
measurement of the inclusive \bsg branching ratio
$\br( \mbsg ) = ( 2.32 \pm 0.57 \pm 0.35) \times 10^{-4}~$, which is in
good agreement with the SM prediction.  
It has been noticed that in a two Higgs doublet model (THDM) the charged
Higgs boson can give a substantial contribution to the \bsg rate
\cite{thdm,GSW}.
In fact, this result constrains the mass of the charged Higgs boson in a
certain type of THDM called Model II \cite{thdm,GSW} to be larger than
$( 244 + 63/(\tan\beta)^{1.3} )$ GeV \cite{cleo} where $\tan\beta$ is
the ratio of the vacuum expectation values of two Higgs fields.

The Higgs sector in the minimal supersymmetric (SUSY) extension of SM 
is a special case of THDM II.
However, the above-mentioned limit cannot be directly applied, because 
SUSY particles can contribute to the \bsg process in addition to the SM
particles and the charged Higgs.
It is a natural question to ask how the charged Higgs mass limit is
modified in the SUSY extension of SM.  
Many authors have discussed the \bsg process in SUSY
\cite{BBMR,Oshimo,Barger,Barbieri,Lopez1,Okada,GaristoNg,Diaz,Borzumati,%
BertoliniVissani,NathArnowitt,Nojiri,Lopez2,Kane,Carena,Kolda}.
In fact, the authors of Ref.~\citen{Barger} considered the constraints on 
the parameter space of the Higgs sector in the minimal supersymmetric
standard model (MSSM) obtained from the \bsg process and noted that this
process is sensitive to the region where the Higgs search in the future
hadron collider turns out to be the most difficult.
It was, however, pointed out that the charged Higgs bounds given by them
cannot be in general valid since they had only taken into account the
charged Higgs effect and neglected the SUSY particle's
contributions \cite{Barbieri}.
Although the SM and the charged Higgs contributions to the \bsg
amplitude have the same sign the SUSY loop can interfere either
constructively or destructively with them and the limit for the charged
Higgs mass from this process can be weakened by the effects of the SUSY
particles.

The minimal supergravity model provides an attractive framework for 
SUSY extension of SM.
In this model, masses and mixing parameters for SUSY particles can be
expressed by a few soft SUSY breaking parameters as well as gauge and
Yukawa coupling constants.
The \bsg branching ratio thus depends on much smaller number of free
parameters compared to that in general SUSY standard models.
It has been noticed \cite{Lopez1,BertoliniVissani,Lopez2} that the sign
of the SUSY loop contributions with respect to those of the SM and
charged Higgs is strongly correlated with the sign of the higgsino mass
parameter \ie the $\mu$ parameter in the minimal supergravity model.

We would like to compare the CLEO data with the prediction of the
minimal supergravity model and determine the allowed region in the
parameter space of the Higgs sector in the model.
Namely, scanning the free parameter space extensively, we search for the 
constraint in the space of the charged Higgs mass and $\tan\beta$.
We study the whole range of $\tan\beta$ which is consistent
with the fact that all of the top, bottom and tau Yukawa coupling
constants remain perturbative up to the grand unification scale. 
Although it is difficult to draw a general conclusion on
the constraint in general SUSY standard models, we can derive
a useful constraint if we restrict ourselves to the case of 
the minimal supergravity model. 
It will be shown that the lower bound of the charged Higgs
mass crucially depends on the sign of $\mu$.
The bound becomes much larger than that in the non-SUSY
THDM II for $\mu<0$, while no strong bound is
obtained for $\mu>0$.

\section{\bsg in the minimal supergravity}

The calculation of the \bsg branching ratio has already been 
discussed extensively in the literature
\cite{GSW,Misiak,BurasMisiakMunzPokorski}.
The decay rate for \bsg normalized to the semileptonic decay rate 
is given by
\begin{eqnarray}
  \frac{\Gamma(\mbsg)}{\Gamma(\ b \rightarrow c e \ol{\nu}\ )}
  &=& \frac{6\alpha_{\rm QED}}{\pi g(m_c/m_b)}
      \frac{|V_{ts}^* V_{tb}|^2}{|V_{cb}|^2}
      \left| C_7^{\rm eff}(Q) \right|^2 ~,
  \label{BRformula}
\\
    C_7^{\rm eff}(Q)
    &=& \eta^{16/23} C_7(M_W)
       + \frac{8}{3}\left(   \eta^{14/23}
                           - \eta^{16/23} \right) C_8(M_W)
       + C ~,
\nonumber
\end{eqnarray}
where $\eta = \alpha_s(M_W)/\alpha_s(Q)$, $Q$ being the scale 
of the order of the bottom mass, 
and $g(z) = 1 - 8z^2 + 8z^6 - z^8 - 24z^4\ln z$.  
$C$ is a constant which depends on $\eta$.
The above formula takes the leading order QCD corrections into account.
The $C_7(M_W)$ and $C_8(M_W)$ are coefficients of the
magnetic and chromomagnetic operators at $M_W$.
The $C$ term is induced by operator mixing in evolving 
from $M_W$ to the low energy scale $Q$.

Ambiguities in the calculation are discussed in detail in
Ref.~\citen{BurasMisiakMunzPokorski}.
The most important ambiguity comes from the choice of the
renormalization scale $Q$.
Varying $Q$ between $m_b/2$ and $2 m_b$ induces an ambiguity
of $\pm 25\ \%$ for the branching ratio in SM.
Other ambiguities include the choice of $m_c/m_b$ (which affects 
the semileptonic rate) and the value of $\alpha_s(M_Z)$.

In MSSM, the coefficients $C_7(M_W)$ and $C_8(M_W)$ receive the
following contributions at one loop:
\begin{enumerate}
\item the $W$ and top quark loop (SM contribution);
\item the charged Higgs and top quark loop;
\item the chargino and up-type squark loops;
\item the gluino and down-type squark loops;
\item \label{ntrlno}the neutralino and down-type squark loops.
\end{enumerate}
The contribution from \ref{ntrlno} is known to be very small
\cite{BBMR}, which we will ignore hereafter.
THDM II amplitude is calculated with use of the first two contributions
only.
The charged Higgs contribution depends on its mass and the
ratio of the vacuum expectation values of the two Higgs doublets,
\ie $\tan\beta =
\vev{H_2^0} / \vev{H_1^0}$, where $H_1^0$ and $H_2^0$ are
the neutral components of the two Higgs doublets.
The chargino and gluino loop contributions depend on the mass
and mixing of the particles inside the loop.
Although the squark mixing matrices are arbitrary parameters 
in a general SUSY standard model, 
they can be calculated from the flavor mixing matrix of quarks 
(the Cabibbo-Kobayashi-Maskawa matrix) in the minimal supergravity model
by solving the renormalization group equations (RGEs) for
various soft SUSY breaking terms.

The soft SUSY breaking parameters at the GUT
scale are the universal scalar mass ($m_0$), a parameter in
the trilinear coupling of scalars ($A_X$), a parameter in
the two Higgs coupling ($B_X$) and the gaugino mass ($M_{gX}$).
We are assuming the GUT relation for the three gaugino masses \ie the
SU(3), SU(2) and U(1) gaugino mass parameters are equal at the GUT
scale.
Besides these soft SUSY breaking parameters, the superpotential contains
the Yukawa coupling constants and the $\mu$ parameter.
Given a set of values for the quark and lepton masses, CKM matrix
elements and $\tan\beta$, we determine all the particle masses and
mixings at the weak scale by solving relevant RGEs with initial
conditions at the GUT scale specified by the above parameters.
We compute the Higgs effective potential at the weak scale and require
that the electroweak symmetry is broken properly (the radiative breaking 
scenario).
We include the one loop corrections to the effective potential induced
by the Yukawa couplings for the third generation.
The condition for radiative breaking with the correct scale reduces the
number of free parameters to three for given $\tan\beta$ and $M_t$.
We can think of these parameters as the charged Higgs mass
($m_{H^\pm}$), SU(2) gaugino mass ($M_2$) and $\mu$ at the weak scale.
For a give set of these five parameters, all other masses and mixings
are calculated.
For the detail of the calculation, see Ref.~\citen{gna-go-gno}.

\section{Numerical results}

We now present the results of the \bsg branching ratio.
Besides the radiative breaking condition we require the following
phenomenological constraints \cite{pdg}:
\begin{enumerate}
\item The mass of any charged SUSY particle is larger than 45 GeV;
\item The sneutrino mass is larger than 41 GeV;
\item The gluino mass is larger than 100 GeV;
\item Neutralino search results at LEP \cite{aleph}, which require
$\Gamma(Z \rightarrow \chi \chi)< 22$ MeV,
$\Gamma(Z \rightarrow \chi \chi')$,
$\Gamma(Z \rightarrow \chi' \chi')< 5 \times 10^{-5}$ GeV,
where $\chi$ is the lightest neutralino and 
$\chi'$ is any neutralino other than the lightest one;
\item The lightest SUSY particle (LSP) is neutral;
\item The condition for not having a charge or color
  symmetry breaking vacuum \cite{aterm}.
\end{enumerate}
\begin{wrapfigure}{l}{7cm}
  \epsfysize=7cm
  \makebox[7cm]{
    \centerline{
      \epsffile{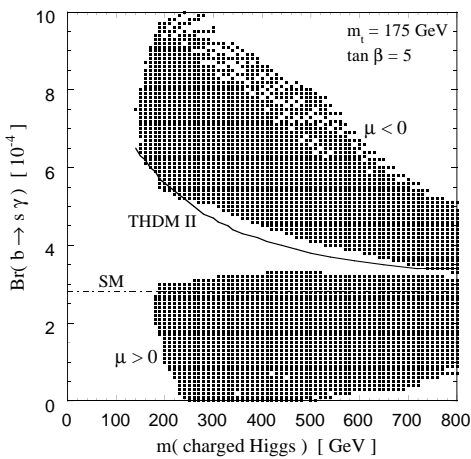}
    }
  }
  \caption{
    \bsg branching ratio for $m_t = 175$ GeV and $\tan\beta = 5$.
    Each dot corresponds to a sample point which satisfies radiative
    breaking and phenomenological constraints (see text).
    Solid line represents the branching ratio calculated with the SM
    and charged Higgs contributions only (THDM II).
    Dot-dashed line represents the SM value.
    }
  \label{fig:br}
\end{wrapfigure}
In Fig.~\ref{fig:br}, we show the \bsg branching ratio for
and $\tan\beta = 5$.
The top quark mass is fixed to $m_t = 175$ GeV \cite{top}%
\footnote{%
This top quark mass is the $\overline{\rm MS}$ running mass
at $Q=M_Z$.
This mass coincides with the pole mass within a few percent
\cite{BurasJaminWeisz}.
}
in the present calculations.
Each point in the figure corresponds to the value of the 
\bsg branching ratio for each scanned point in the parameter
space compatible with the above conditions.
This branching ratio includes the chargino and gluino loop
contributions as well as the SM and the charged Higgs loop. 
The line in the figure represents the branching ratio when 
only the SM and charged Higgs contributions are retained.
We notice that the points are divided by this line.
In fact, the points above and below this line correspond
to $\mu < 0$ and $\mu > 0$ cases respectively%
\footnote{
Our convention of the sign of $\mu$ is opposite to those 
in Refs.~\citen{BertoliniVissani} and \citen{Lopez2}.
}.
This confirms the assertion \cite{Lopez1,BertoliniVissani,Lopez2} that 
the sign of $\mu$ determines whether the SUSY contribution enhances or
suppresses the \bsg branching ratio.

\begin{figure}
  \makebox[7cm]{
    \epsfysize=6cm
    \centerline{
      \epsffile{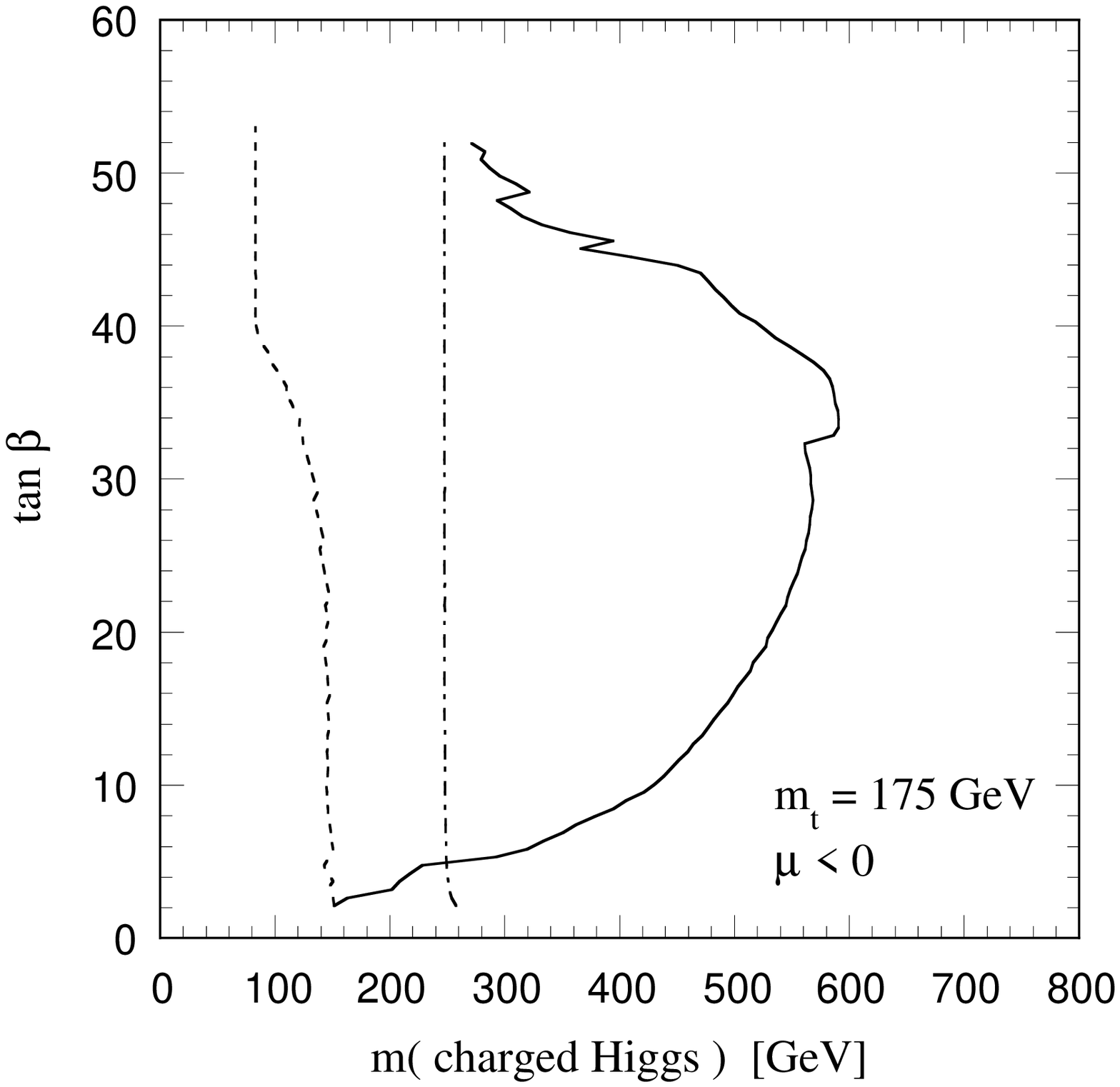}
      }
    }
  \makebox[7cm]{
    \epsfysize=6cm
    \centerline{
      \epsffile{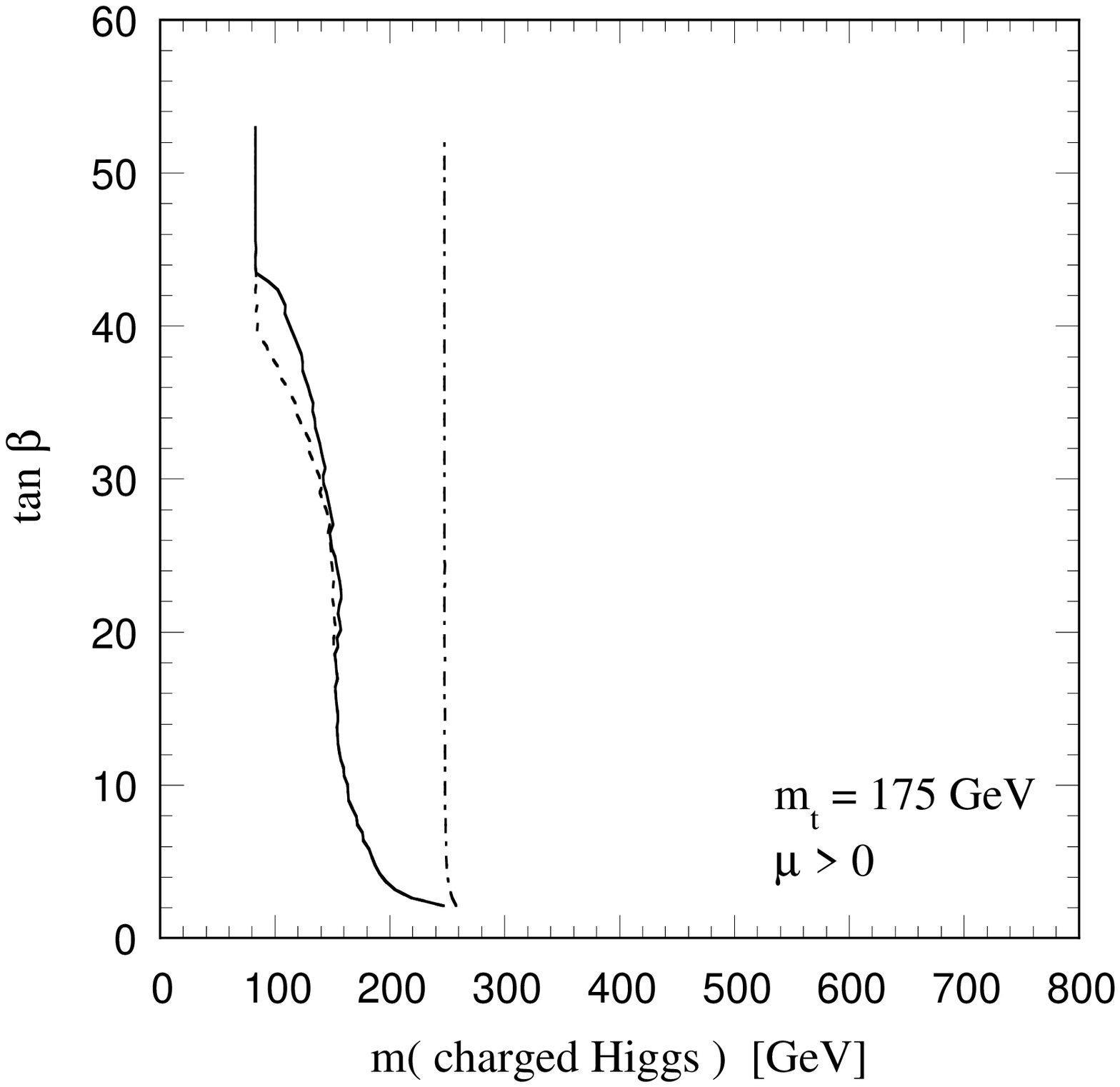}
      }
    }
  \caption{
    Excluded region in the $\tan\beta$ and $m_{H^\pm}$ space for
    either sign of $\mu$.
    Each line represents the lower bound for the charged Higgs mass;
    solid line: all constraints included; dashed line: without \bsg
    constraint; dot-dashed line: THDM II with \bsg constraint.
    }
  \label{fig:exclude}
\end{figure}

We show the excluded region in the $\tan\beta$ and $m_{H^\pm}$ space 
in Fig.~\ref{fig:exclude}.
The range of the $\tan\beta$ we have scanned are $2\lsim \tan\beta \lsim
55$. For the values of $\tan\beta$ larger or smaller than this
range the Yukawa coupling constant for top or bottom/tau
blows up below the GUT scale.
The two branches $\mu>0$ and $\mu<0$ are separately plotted. 
The  excluded region is determined using the CLEO
result  $1 \times 10^{-4} < \br(\mbsg) < 4 \times 10^{-4}$.
In order to take account of the theoretical uncertainties
we have calculated the \bsg branching ratio by varying the
renormalization scale $Q$ between $m_b/2$ and $2 m_b$.
There are other sources of theoretical ambiguities
which are expected to be minor compared to the choice
of the renormalization scale. These includes the choice of
the $m_c/m_b$, $\alpha_s(m_Z)$, CKM matrix elements, etc.
In the standard model, the unitarity of the CKM matrix and the smallness 
of the quantity $V_{us}^* V_{ub}$ guarantees 
the \bsg amplitude to be proportional to a single factor
$V_{ts}^* V_{tb} \approx -V_{cs}^* V_{cb}$, so that the uncertainty from 
the CKM matrix element is small.
The same situation occurs in the minimal supergravity model where the
flavor mixing is essentially determined by the Yukawa couplings.
Consequently the uncertainty from the choice of CKM matrix elements is
small in our case.
In the calculation we have used $ \alpha_s(m_Z)=0.116$,
$m_c(m_c)=1.27$ GeV, $m_b(m_b)=4.25$ GeV, $|V_{us}|=0.221$,
$|V_{cb}|=0.041$, $|V_{ub}|/|V_{cb}|=0.08$ and $\delta_{13}=\pi/3$ where
$\delta_{13}$ is the CP phase with the convention used in
Ref.~\citen{pdg}.
Taking into account of these ambiguities other than the renormalization
scale dependence, we allow an additional 10\% uncertainty in order to
obtain a conservative bound.
We regard a point in $(\ \tan\beta,\ m_{H^\pm}\ )$ space
excluded when the branching ratio cannot be within the CLEO
bound for any choice of soft SUSY breaking parameters even if
we consider the above-mentioned theoretical ambiguities.
We also show in these figures the lower bound of the charged
Higgs mass when only the SM and the charged Higgs contributions
are retained.
In comparison, the excluded region in minimal supergravity becomes
larger when $\mu < 0$. The bound reaches 400 GeV for $\tan \beta > 10$%
\footnote{
  If we reduce the theoretical uncertainty by fixing the renormalization 
  scale $Q=m_b$, the lower bound of $m_{H^\pm}$ for $\mu<0$ is raised by 
  $\sim$ 100 GeV.
}.
For $\mu > 0$, the \bsg process is not very effective in constraining
the charged Higgs mass, because the charged Higgs contribution can be
completely cancelled by SUSY particle contributions.
It is also interesting to see the $\tan\beta$ dependence of the charged
Higgs mass bound for $\mu < 0$. The bound becomes strongest around
$\tan\beta\sim 35$ and becomes weaker for larger values of $\tan\beta$.   

The reason for the strong dependence on $\mu$ can be understood as
follows. For the chargino and up-type squark loop the most important
contribution comes from the loop diagram with the top and bottom Yukawa
coupling constants. This diagram is proportional to a product of the
left-right mixing parameter of the stops, \ie $A_t + \mu\cot\beta$, and
the higgsino mixing parameter $\mu$.
The parameter $A_t$ is determined by $A_X$ and $M_{gX}$ through the
RGEs.
An interesting observation is that for such a high value of the top
quark mass as considered here $A_t$ is almost independent of $A_X$ and
proportional to $M_{gX}$.
Moreover, the $\mu\cot\beta$ term is suppressed for  $\tan\beta > 3$.
Therefore, the amplitude from this sector is proportional to a product
of the gaugino mass and the higgsino mass parameter in a very good
approximation.
A similar consideration applies to the gluino and down-type squark loop.
In this case a sizable contribution can arise when the graph
involves the left-right mixing in the sbottom sector, especially
for a large value of $\tan\beta$. Then, the amplitude is proportional
to a product of the gluino mass $(M_3)$ and the sbottom mixing
parameter, \ie  $A_b + \mu\tan\beta$. For a large value of $\tan\beta$
the latter parameter is governed by the $\mu \tan \beta$ term.
Here again, the contribution to the amplitude is almost 
proportional to the product of the gaugino mass and the higgsino 
mass parameter. In both cases, the SUSY
contribution enhances (suppresses) the SM amplitude when
$\mu < 0$ ($\mu > 0$). This will explain the tendency seen
in Fig.\ \ref{fig:br}\footnote{The reason for the strong 
dependence on the sign of $\mu$ is analyzed for example
in Ref.~\citen{Kolda} where they concentrated the case
of small and very large values of $\tan \beta$ motivated
by the Yukawa unification.}. 
\begin{figure}
  \makebox[6cm]{
    \epsfysize=6cm
    \centerline{
      \epsffile{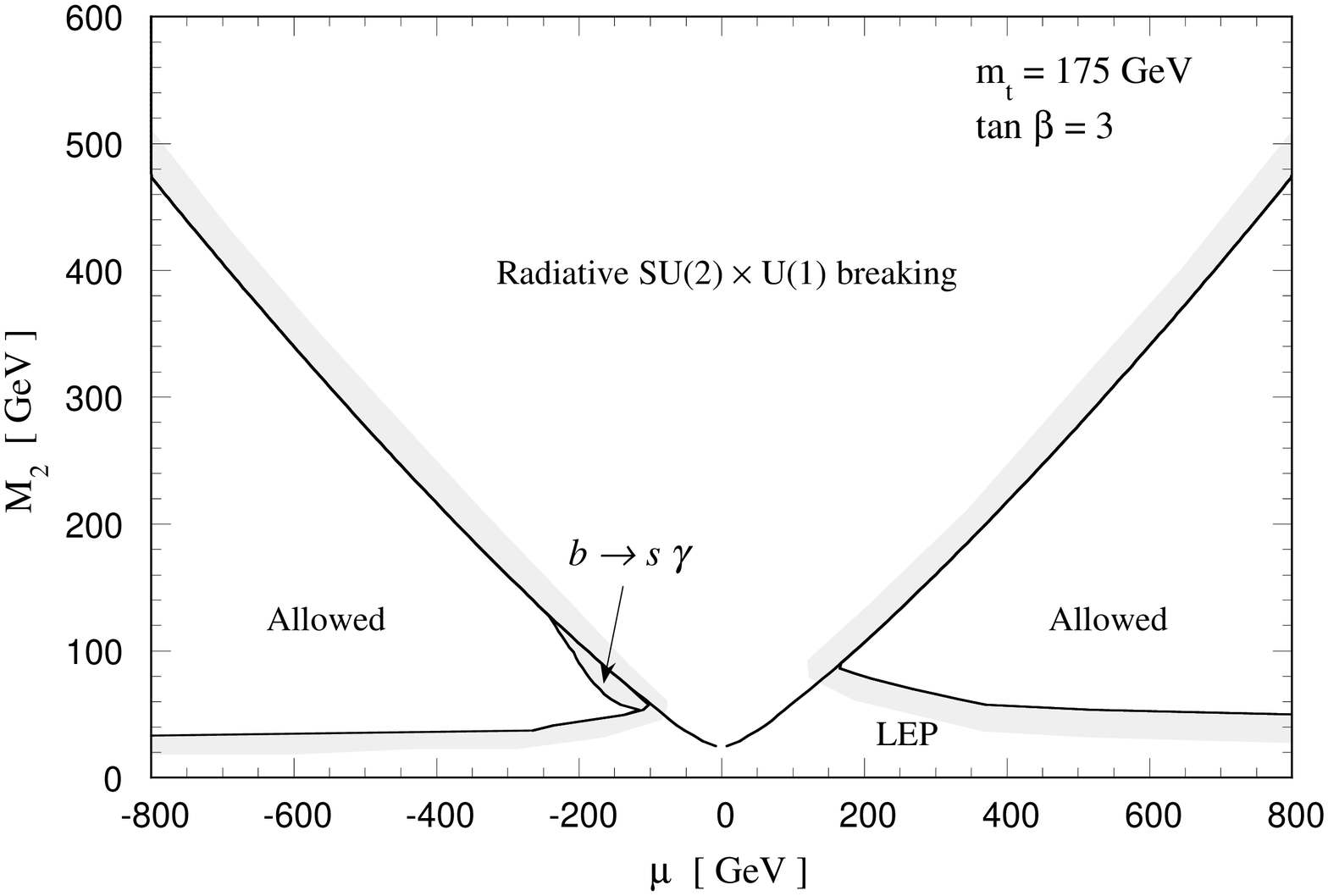}
      }
    }
  \makebox[10cm]{
    \epsfysize=6cm
    \centerline{
      \epsffile{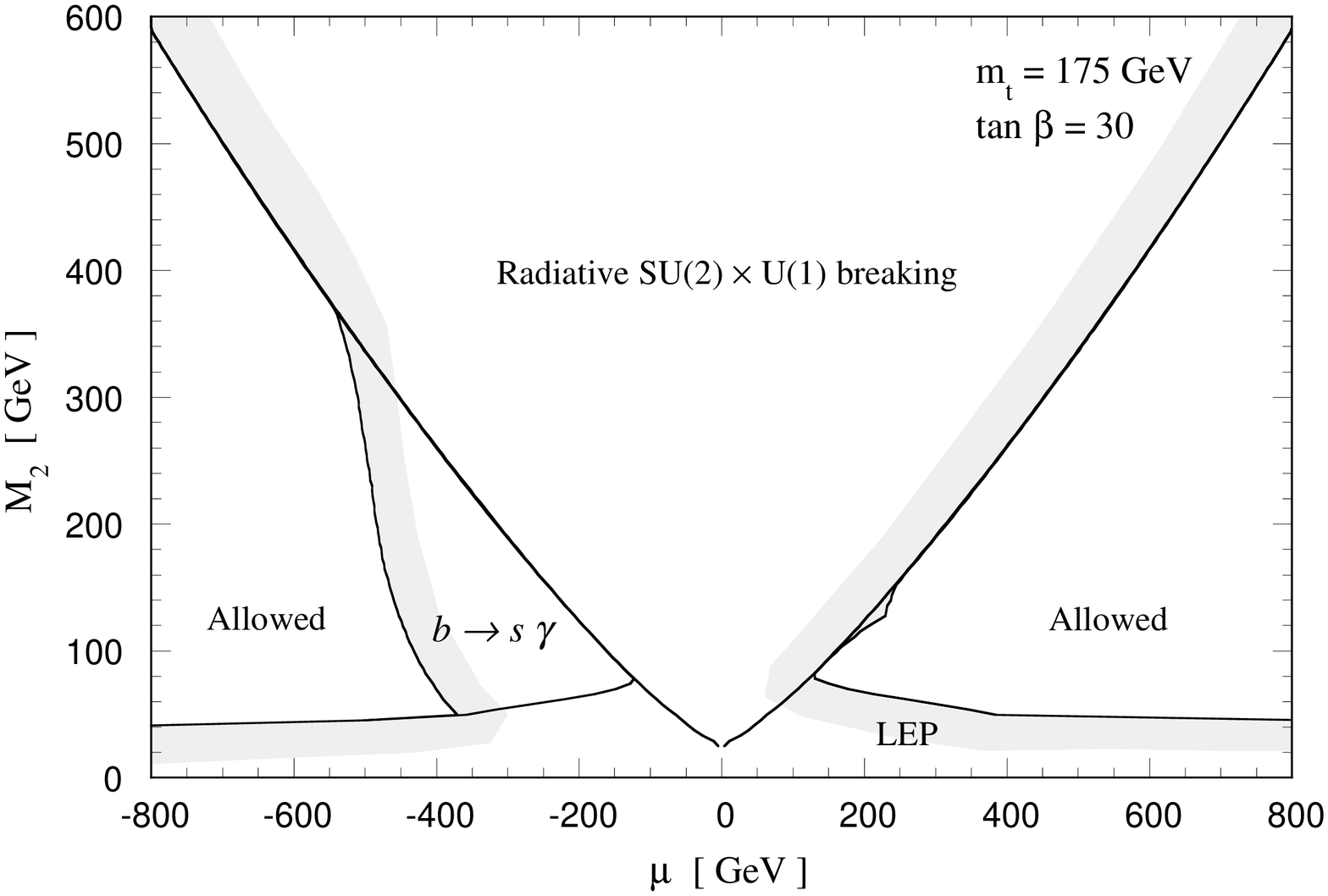}
      }
    }
  \caption{
    Allowed region in $\mu$--$M_2$ space for $\tan\beta =$ 3 and 30.
    }
  \label{fig:muM2}
\end{figure}

The lower bound of the charged Higgs mass for $\mu < 0$ is determined by
the minimum value of the branching ratio for a fixed choice of
$m_{H^{\pm}}$ and $\tan \beta$.
From the above discussion the chargino and stop loop contribution is
approximately proportional to the product of the top and bottom Yukawa
coupling constants, namely $\frac{m_b m_t}{\sin{\beta} \cos{\beta}}$.
Also the branching ratio depends on the mass of stop/chargino.
To determine the minimal contribution we need to know the maximum value
of the stop/chargino mass for a given set of $m_{H^\pm}$ and
$\tan\beta$.
For each given set, we can scan the remaining two parameters ($\mu$,
$M_2$) allowed by the phenomenological constraints and the condition for
the electroweak radiative breaking (see Fig.~\ref{fig:muM2}).
In general the allowed region becomes a strip for each sign of $\mu$. 
Up to $\tan \beta \sim 35 $ the maximum value of the lighter stop mass
relative to the charged Higgs mass does not strongly depends on
the value of $\tan \beta$.
Then the $\tan \beta$ dependence of the minimum branching ratio is
essentially determined by $\frac{m_b m_t}{\sin{\beta}\cos{\beta}}$.
Close investigations show that beyond $\tan \beta \sim 35 $ where the
bottom and tau Yukawa coupling constants becomes as large as the top
Yukawa coupling constant the condition for the radiative symmetry
breaking allows the larger values of the stop mass, therefore the
contribution from the chargino and stop loop can be much smaller.
This explains why the lower bound becomes smaller for the value
of $\tan \beta > 35$.
Note that the lower bound of the charged Higgs mass exceeds that of the
Model II THDM even in the case of  $\tan \beta > 35$ since the chargino
and stop loop effect does not change its sign.
In fact, the bound approaches the line of the THDM II for the maximal
value of $\tan \beta$.

\section{Conclusion}

To summarize, we have shown that for $\mu<0$ the lower bound of the
charged Higgs mass increases compared to that in the THDM II.
On the other hand, for  $\mu>0$, the \bsg process cannot put useful
constraints on the allowed range of the charged Higgs mass because it is
possible that the charged Higgs contribution is completely cancelled by
other SUSY contributions.
We have also pointed out that a parameter region corresponding to the
charged Higgs mass less than 180 GeV and $3 \lsim \tan\beta \lsim 5$ is
excluded by the \bsg process irrespective of the sign of $\mu$.
The constraints obtained here are important for the SUSY Higgs search in
the future colliders.
For example the parameter region in the ($m_{H^{\pm}}$,$\tan{\beta}$)
space where no signal could be found at LHC roughly corresponds to the
region near 150 GeV $\lsim m_{H^{\pm}}\lsim$ 300 GeV, $5 \lsim \tan
\beta \lsim 10$ although details of the Higgs search limits depend on
other parameters like the stop mass or expected detector performance,
etc. \cite{atlas}.
It is interesting to see that most of these region are already excluded
for  $\mu<0$. On the other hand if LHC or linear colliders find the
MSSM Higgs and investigation of its property shows that the Higgs
parameters are, say, $\tan {\beta}\sim 10 - 20$ and charged Higgs mass
$\sim$ 300 GeV, we can conclude that the $\mu$ should be positive since
otherwise the $b\rightarrow s \gamma$ constraint cannot be satisfied.

\end{document}